\documentclass{iopart}

\usepackage{graphicx}
\usepackage{subfig}
\usepackage{multirow}
\usepackage{tabu}

\newcommand{\be}{\begin{equation}}
\newcommand{\ee}{\end{equation}}
\newcommand{\bea}{\begin{eqnarray}}
\newcommand{\eea}{\end{eqnarray}}

\begin{document}

\title{Constraints on LISA Pathfinder's self-gravity: design requirements, estimates  and testing procedures}

\author{
M~Armano$^{a}$,
H~Audley$^{b}$,
G~Auger$^{c}$,
J~Baird$^{n}$,
P~Binetruy$^{c}$,
M~Born$^{b}$,
D~Bortoluzzi$^{d}$,
N~Brandt$^{e}$,
A~Bursi$^{t}$,
M~Caleno$^{f}$,
A~Cavalleri$^{u}$,
A~Cesarini$^{g}$,
M~Cruise$^{h}$,
K~Danzmann$^{b}$, 
M~de Deus Silva$^{a}$,
D~Desiderio$^{t}$,
E~Piersanti$^{t}$,
I~Diepholz$^{b}$, 
R~Dolesi$^{g}$,
N~Dunbar$^{i}$,
L~Ferraioli$^{j}$,
V~Ferroni\footnote{corresponding author: valerio.ferroni@unitn.it}$^{g}$,
E~Fitzsimons$^{e}$,
R~Flatscher$^{e}$,
M~Freschi$^{a}$,
J~Gallegos$^{a}$,
C~Garc\'ia Marirrodriga$^{f}$,
R~Gerndt$^{e}$,
L~Gesa$^{k}$,
F~Gibert$^{g}$,
D~Giardini$^{j}$,
R~Giusteri$^{g}$,
C~Grimani$^{l}$,
J~Grzymisch$^{f}$,
I~Harrison$^{m}$,
G~Heinzel$^{b}$, 
M~Hewitson$^{b}$, 
D~Hollington$^{n}$,
M~Hueller$^{g}$,
J~Huesler$^{f}$,
H~Inchausp\'e$^{c}$,
O~Jennrich$^{f}$,
P~Jetzer$^{o}$,
B~Johlander$^{f}$,
N~Karnesis$^{b}$,
B~Kaune$^{b}$,
N~Korsakova$^{b}$,
C~Killow$^{p}$,
I~Lloro$^{k}$,
L~Liu$^{g}$,
R~Maarschalkerweerd$^{m}$,
S~Madden$^{f}$,
D~Mance$^{j}$,
V~Mart\'{i}n$^{k}$,
L~Martin-Polo$^{a}$,
J~Martino$^{c}$, 
F~Martin-Porqueras$^{a}$,
I~Mateos$^{k}$,
P~McNamara$^{f}$,
J~Mendes$^{m}$,
L~Mendes$^{a}$,
A~Moroni$^{t}$,
M~Nofrarias$^{k}$,
S~Paczkowski$^{b}$,
M~Perreur-Lloyd$^{p}$,
A~Petiteau$^{c}$, 
P~Pivato$^{g}$,
E~Plagnol$^{c}$, 
P~Prat$^{c}$,
U~Ragnit$^{f}$,
J~Ramos-Castro$^{q}$$^{r}$,
J~Reiche$^{b}$,
J~A~Romera Perez$^{f}$,
D~Robertson$^{p}$, 
H~Rozemeijer$^{f}$,
F~Rivas$^{k}$,
G~Russano$^{g}$,
P~Sarra$^{t}$,
A~Schleicher$^{e}$,
J~Slutsky$^{s}$,
C~Sopuerta$^{k}$,
T~Sumner$^{n}$, 
D~Texier$^{a}$,
J~Thorpe$^{s}$,
R~Tomlinson$^{i}$
C~Trenkel$^{i}$,
D~Vetrugno$^{g}$,
S~Vitale$^{g}$,
G~Wanner$^{b}$, 
H~Ward$^{p}$,
P~Wass$^{n}$, 
D~Wealthy$^{i}$,
W~Weber$^{g}$,
A~Wittchen$^{b}$,
C~Zanoni$^{d}$,
T~Ziegler$^{e}$,
P~Zweifel$^{j}$}

\address{$^{a}$ European Space Astronomy Centre, European Space Agency, Villanueva de la
Ca\~{n}ada, 28692 Madrid, Spain}
\address{$^{b}$ Albert-Einstein-Institut, Max-Planck-Institut f\"ur
Gravitationsphysik und Leibniz Universit\"at  Hannover, 30167 Hannover, Germany}
\address{$^{c}$ APC UMR7164, Universit\'e Paris Diderot, Paris, France}
\address{$^{d}$ Department of Industrial Engineering, University of Trento, via Sommarive 9, 38123 Trento, 
and Trento Institute for Fundamental Physics and Application / INFN}
\address{$^{e}$ Airbus Defence and Space, Claude-Dornier-Strasse, 88090 Immenstaad, Germany}
\address{$^{f}$ European Space Technology Centre, European Space Agency, 
Keplerlaan 1, 2200 AG Noordwijk, The Netherlands}
\address{$^{g}$ Dipartimento di Fisica, Universit\`a di Trento and Trento Institute for 
Fundamental Physics and Application / INFN, 38123 Povo, Trento, Italy}
\address{$^{h}$ Department of Physics and Astronomy, University of
Birmingham, Birmingham, UK}
\address{$^{i}$ Airbus Defence and Space, Gunnels Wood Road, Stevenage, Hertfordshire, SG1 2AS, UK }
\address{$^{j}$ Institut f\"ur Geophysik, ETH Z\"urich, Sonneggstrasse 5, CH-8092, Z\"urich, Switzerland}
\address{$^{k}$ Institut de Ci\`encies de l'Espai (CSIC-IEEC), Campus UAB, Carrer de Can Magrans s/n, 08193 Cerdanyola del Vall\`es, Spain}
\address{$^{l}$ DiSBeF, Universit\`a di Urbino "Carlo Bo", Via S. Chiara, 27 61029 Urbino/INFN, Italy}
\address{$^{m}$ European Space Operations Centre, European Space Agency, 64293 Darmstadt, Germany }
\address{$^{n}$ The Blackett Laboratory, Imperial College London, UK}
\address{$^{o}$ Physik Institut, 
Universit\"at Z\"urich, Winterthurerstrasse 190, CH-8057 Z\"urich, Switzerland}
\address{$^{p}$ SUPA, Institute for Gravitational Research, School of Physics and Astronomy, University of Glasgow, Glasgow, G12 8QQ, UK}
\address{$^{q}$ Department d'Enginyeria Electr\`onica, Universitat Polit\`ecnica de Catalunya,  08034 Barcelona, Spain}
\address{$^{r}$ Institut d'Estudis Espacials de Catalunya (IEEC), C/ Gran Capit\`a 2-4, 08034 Barcelona, Spain}
\address{$^{s}$ NASA Goddard Space Flight Center, 8800 Greenbelt Road, Greenbelt, MD 20771, USA}
\address{$^{t}$ CGS S.p.A, Compagnia Generale per lo Spazio, Via Gallarate, 150 - 20151 Milano, Italy}
\address{$^{u}$ Istituto di Fotonica e Nanotecnologie, CNR-Fondazione Bruno Kessler, I-38123 Povo, Trento, Italy}

\begin{abstract}
LISA Pathfinder satellite has been launched on 3th December 2015 toward the Sun-Earth first Lagrangian point (L1) where the LISA Technology Package (LTP), which is the main science payload, will be tested. With its cutting-edge technology, the LTP will provide the ability to achieve unprecedented geodesic motion residual acceleration measurements down to the order of $3 \times 10^{-14}\,\mathrm{m/s^2/{Hz^{1/2}}}$ within the $1-30\,\mathrm{mHz}$ frequency band.
The presence of the spacecraft itself is responsible of the local gravitational field which will interact with the two proof test-masses. 
Potentially, such a force interaction might prevent to achieve the targeted free-fall level originating a significant source of noise. We balanced this gravitational force with sub $\mathrm{nm/s^2}$ accuracy, guided by a protocol based on measurements of the position and the mass of all parts that constitute the satellite, via finite element calculation tool estimates. In the following, we will introduce requirements, design and foreseen on-orbit testing procedures. 
\end{abstract}



\section{Introduction}
\label{s1}

\begin{figure}[htbp]
\begin{center}
\subfloat[TM interferometric link]{\includegraphics[width=150pt]{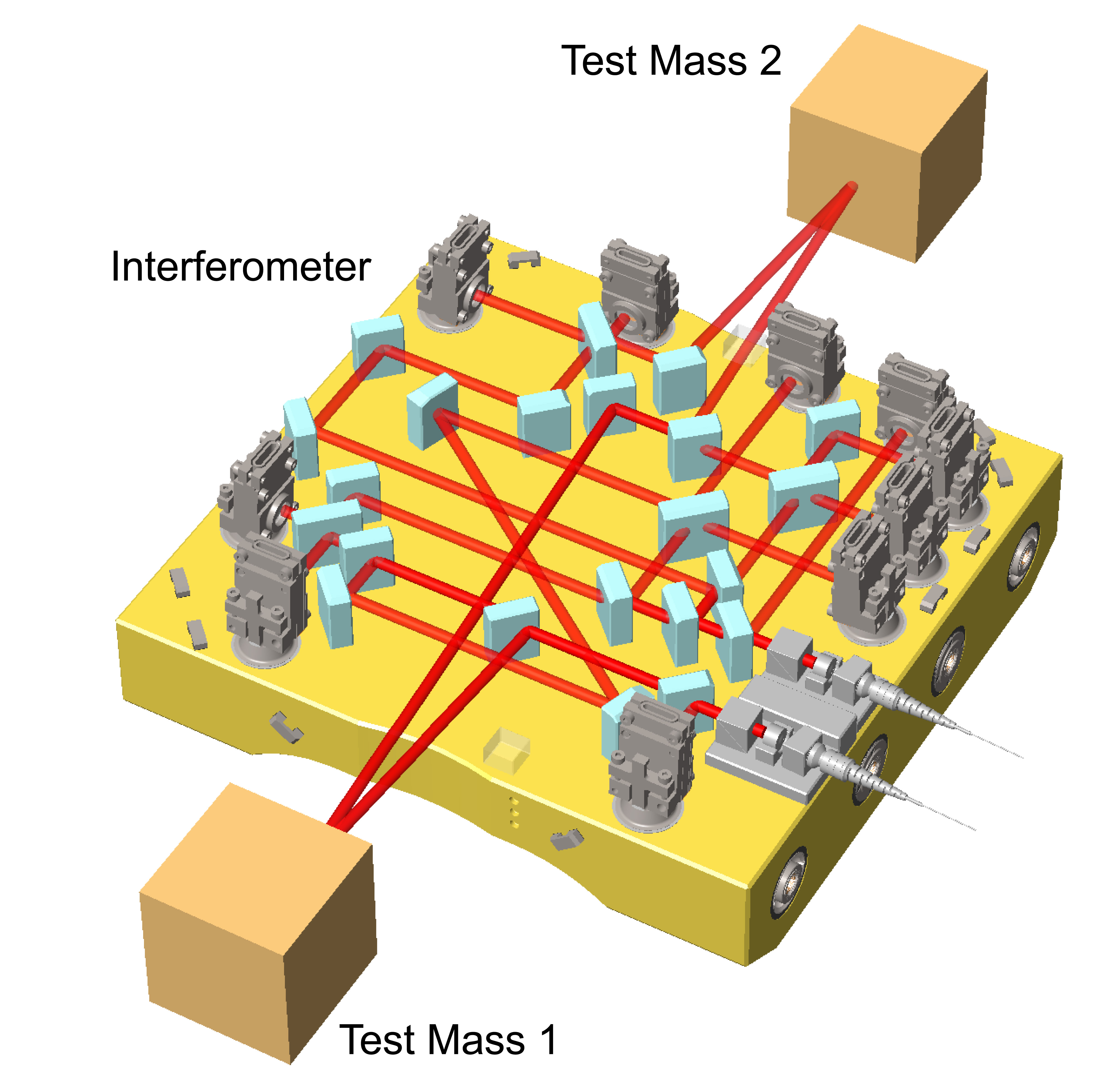}}%
\qquad
\subfloat[TM reference axes]{\includegraphics[width=150pt]{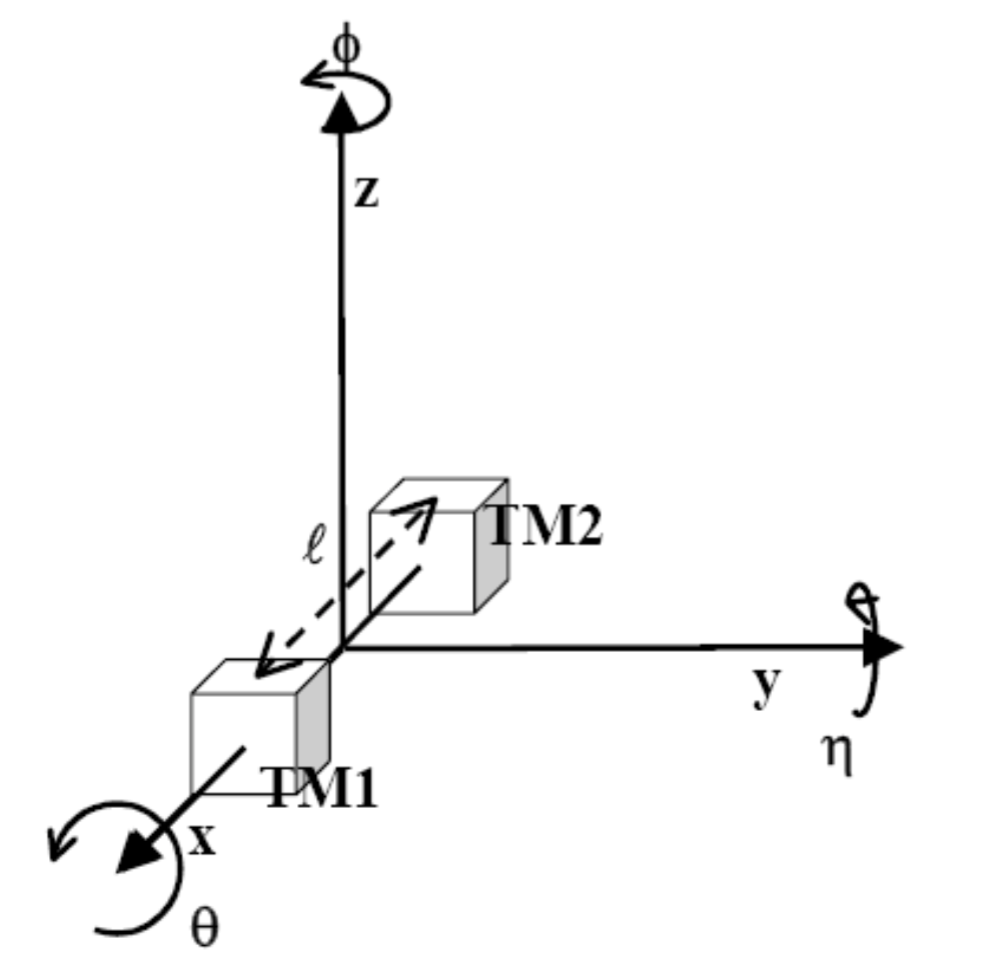} }%
\caption{LTP conceptual picture}
\label{f0}
\end{center}
\end{figure}

Shortly after commissioning phase of Lisa Pathfinder (LPF) \cite{lpf,LPFrequirements}, the LISA Technology Package (LTP) is going to be calibrated and probed with a battery of on-flight experiments. The LISA Pathfinder differential accelerometer will orbit around
the Sun-Earth Lagrangian point L1.  In addition to allowing a stable
orbit in the Sun-Earth reference frame, this provides a
background gravitational gradient seven orders of magnitude smaller than
the low earth orbit occupied by GOCE and other
proposed geodesy differential accelerometers. Hence, very soon, the LPF scientific payload, which is sketched in Figure \ref{f0}, will perform the most sensitive
measurements of differential acceleration between free falling test bodies using the first space high precision interferometer link, ever flown on-board of an orbiting satellite.
By measuring the differential acceleration between two test masses (TMs), one of them being left free floating, (so nominally following a pure geodesic), and acting as the reference of the measurement, LPF will verify whether a drag-free proof TM can be kept along one measuring axis, x axis in Figure \ref{f0}, in pure gravitational free fall having a residual acceleration noise:
\be
\label{req}
\fl \qquad S_a^{1/2}(f)\leq 3\times 10^{-14}\left[1 + \left(\frac{f}{3~\textup{mHz}}\right)\right]\frac{\textup{m}}{\textup{s}^2}\frac{1}{\sqrt{\textup{Hz}}}\;;\quad 1~\textup{mHz}\leq f\leq 30~\textup{mHz}\;.
\ee

Fulfilling this requirement on the acceleration represents an important step beyond the current limit \cite{Goce,MiPe} and a step closer, factor of ten, for reaching the mission pre-requisite to build space-born gravitational wave observatories \cite{LISA,eLISA}. 

The presence of gravitational interactions between satellite and the proof TMs modifies the local geodesic and the two TMs won't follow the same geodesic because of non-uniformity of the local gravitational field.
This residual field needs to be balanced by means of electrostatic actuation forces\footnote{In such a context, the Drag-free attitude control subsystem (DFACS) will provide the ability to keep the TM centered.}, that invariably fluctuate themselves due to voltage instabilities. This adds noise to the TM motion in the frequency band of interest:
\be
\label{sadc}
\fl \qquad S_{a,DC}^{1/2} =2\,\lambda\,{\Delta a_x} S^{1/2}_{dV/V}= 7.8\times 10^{-15}\left(\frac{\lambda}{1}\right)\left(\frac{\Delta a_x}{0.65\,\mathrm{\frac{nm}{s^2}}}\right)\left(\frac{S^{1/2}_{dV/V}}{\frac{6\times 10^{-6}}{\sqrt{\mathrm{{Hz}}}}}\right)\frac{\mathrm{m}}{\mathrm{s^2}} \frac{1}{\sqrt{\mathrm{Hz}}}\;,\quad
\ee
where $S_{\delta V/V}^{1/2}$ is the relative amplitude stability of the applied actuation voltages and $\lambda$ is an order-unity factor dependent on the possible correlation the four amplifiers used for $x$ and $\phi$ actuation [7]. $S_{\delta V/V}^{1/2}$ has been measured on ground to be between 3 and 8 ppm/Hz$^{1/2}$ for the relevant amplifiers to be used in flight.  A DC gravitational imbalance of 0.65~nm/s$^2$ will thus produce a relative acceleration noise of 7.8~fm/s$^2$/Hz$^{1/2}$, and thus comfortably within the mission requirements in eqn. \ref{req}.

On the other hand, eqn. \ref{sadc} shows that if the difference in gravitational force per unit mass between TMs exceeds $2.5~\mathrm{n m {s}^{-2}}$, such an imbalance alone - i.e. in the absence of any other noise - would make it impossible to achieve the mission goal, eqn. \ref{req}.
Thus the only way this mission succeeds is to balance $\Delta a_x$ to nearly zero. This is a crucial point for LPF and is the main driver of building the entire plan of gravitational control for this mission, which will be described in this paper.

Other components $\Delta a_y$ and $\Delta a_z$ of the  imbalance, and TM angular acceleration, substitute $\Delta a_x$ in eqn. \ref{sadc} through cross-talk coefficients, and feed into noise on the TM free fall along x.  In addition self-gravity also produces through its force gradient, a parasitic coupling between the spacecraft (S/C) and the TMs that multiplies the read-out noise and the S/C jitter and adds-up to the other disturbances.

To control the noise budget of the mission and to specify the systemic parameters as electrode-to-TM gaps, voltage fluctuations, and cross-talk coefficients, we set a number of requirements on the TM imbalance, the TM unit of mass gravitational force, the TM unit of inertia gravitational torque, and their gradients with respect to the TM coordinates, the LPF gravitational parameters.

The main requirement applies to $\Delta {\vec{a}}$, the imbalance between the TMs \footnote{In practice the gravitational imbalance plus smaller non-gravitational contributions.}; it is expressed, the pedix $_r$ indicating requirement, by the inequalities:
\be
\label{reqdc}
\fl \qquad \left|\Delta {a_x}\right|_{r} \leq 0.650\,\mathrm{nm/s^2}\;;\; \left| \Delta {a_y}\right|_{r} \leq 1.100\,\mathrm{nm/s^2}\;; \;  \left|\Delta {a_z}\right|_{r} \leq 1.850\,\mathrm{nm/s^2}\;.\;\;
\ee

The first goal of the gravitational control plan is fulfilling inequalities \ref{reqdc}.

The requirements on the other gravitational parameters are subordinate to inequalities \ref{reqdc}, were dealt with mostly in the design, and taken under strict control during the system assemblage. They are all listed in the Appendix.

\section{LPF\label{s2}}

The LPF TMs are each placed inside an inertial sensor (IS) which monitors and controls their position and attitude with respect to the spacecraft using a set of capacitors. 
An high precision interferometer provides the laser link along the axis (x axis in Figure \ref{f0}) joining the two TMs for direct measuring the distance of the nominal drag-free reference (TM1) with respect to the S/C and the TMs relative distance, and supplementary angular measurements about the axes, y and z. 
Each of the IS is enclosed inside the inertial sensor heading (ISH) with two important lock mechanisms: the first one, the grabbing, positioning, release mechanism (GPRM) is to hold the TMs before its final release into free-fall; the second, the caging and venting mechanism (CVM), has the double function of constraining the TMs until LPF reaches the desired orbit and of venting to space to provide the ultra-vacuum inside the ISH. 

The two ISHs and the interferometer constitute LTP.  The LTP core assembly (LCA), placed inside the spacecraft's inner cylinder, consists of the following main subsystems: 2 Inertial Sensor Heads (ISH),  the Venting Ducts, installed at the bottom part of the ISHs, the Optical Bench Assembly (OBA), the LCA Assembly Integration Equipment (AIE), and the Diagnostic instruments (DDS) included inside S/C inner cylinder where the LCA is placed. 

\begin{figure}[htbp]
\begin{center}
\subfloat[LTP Core Assembly (LCA)]{\includegraphics[width=230pt]{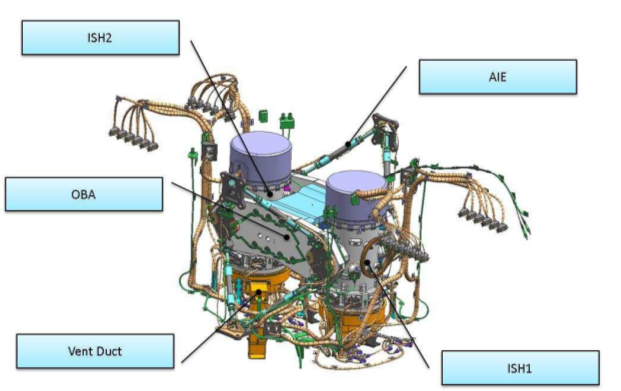}}
\qquad
\subfloat[LPF satellite]{\includegraphics[width=180pt]{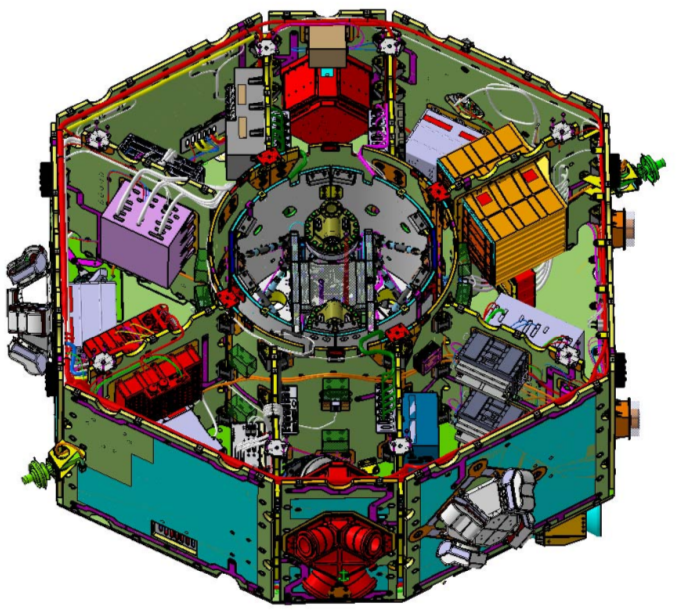}}
\caption{LTP and S/C hardware. }
\label{f1f2}
\end{center}
\end{figure}

The LPF spacecraft will carry on-board another science payload, the Disturbance and Reduction System (DRS) module \cite{DRS}, a scientific contribution from NASA. DRS uses the sensor information supplied by LTP and a set of proprietary micro thrusters to produce a complimentary test of the free fall.
The design (and so the gravitational modeling) and the manufacturing of the LCA was under the responsibility of CGS, with the support of University of Trento, while Airbus (formerly Astrium) managed the DRS and all the remaining parts of the satellite. Figure \ref{f1f2} shows the CAD models of LCA and of the entire S/C.

\section{Spacecraft self-gravity modeling}
\label{s3}

\begin{figure}[htbp]
\begin{center}
\includegraphics[width=150pt]{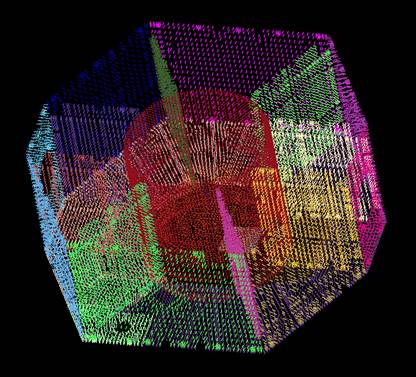}
\caption{SC structure finite element model.}
\label{f3}
\end{center}
\end{figure}

The gravitational field on the TMs was calculated on the basis of the CAD model of LPF;  starting from the smallest screw - for example $1$ g at $10$ cm gives you an acceleration (on the near TM) of  $0.007$ nm/s$^{2}$ - each single piece of the satellite was discretised as a set of point masses, our gravitational mathematical model, GMM. Shown in Figure \ref{f3} is the finite element model of the SC structure. Each point mass individual contribution to the TM acceleration is expressed by the formulas of the gravitational acceleration of an homogeneous parallelepiped (dimensions $L_x,L_y,L_z$) due to the point-like source (of mass $m_s$):
\be
\label{gravpotint}
\fl \qquad\vec{A} = \frac{Gm_s}{L_xL_yL_z}\nabla_{x,y,z}\int_{-Lx}^{Lx}\int_{-Ly}^{Ly}\int_{-Lz}^{Lz}
\frac{dx'dy'dz'}{\sqrt{(x'-x)^2+(y'-y)^2+(z'-z)^2}}\;.
\ee
$\nabla_{x,y,z}$ operates the gradient with respect to the source coordinates, the unprimed ones; the primed coordinates indicate the position of a mass element $dm$ \footnote{$dm=\rho \;dx'dy'dz'= \frac{M}{L_xL_yL_z}dx'dy'dz'$.} of the parallelepiped. The formulas of the single gravitational parameters were derived analytically as closed form expressions in terms of elementary functions from the general expression given in eqn. \ref{gravpotint}. Some details of this not difficult calculation, which uses tabulate integrals, are found in \cite{Armano}.  
 
The gravitational parameter values were firstly calculated by using the CAD/CAM model. Subsequently, these estimates were updated with the as-built position, alignment and mass of the real hardware, measured and acquired during the extensive assembly / integration / test (AIT) campaign. \footnote{In practice the AIT measurements needed to be complaint to specific accuracy to avoid that the errors propagating spoiled the error budget. To set that accuracy, that is the tolerances in the AIT measurements, we performed during the design phase an extensive sensitivity analysis.}.
The measurements were loaded from a report where every single mounted object is recorded together with its mass and positioning info including any defect and discrepancy with respect to the CAD design.  More than ten thousand weighings and recordings were performed and integrated in the calculations. 

The mass measurements lead to a refinement of the mass density value $\rho_{meas}$. On the assumption that the bodies were isotropic, we recalculated the GMMs by multiplying for the normalized measured density $\rho_{meas}/\rho_0$, $\rho_0$ being the nominal density.
The GMMs were recalculated again using the positioning and alignment measurements, translating and rotating the nominal GMMs. 

The gravitational parameters were calculated from these refined GMMs. The entire process was done automatically by software once the measurements became available. Depending on what measurements have been fed, the gravitational parameters changed their refinement level starting from level {\it A}, to the final fifth level {\it E}, through four intermediate steps, where items' mass, items' position, subsystems' mass, and finally subsystems' positions were measured and implemented.      
Joint reviews with the personnel working on both LTP and S/C gravitational parameters occurred periodically while the refinement level passed from A to E.

\section{Gravitational compensated imbalance}
\label{s4}
\begin{figure}[htbp]
\begin{center}
\subfloat[Location. Not seen, inside the Electrode housing, the TM.]{\includegraphics[width=150pt]{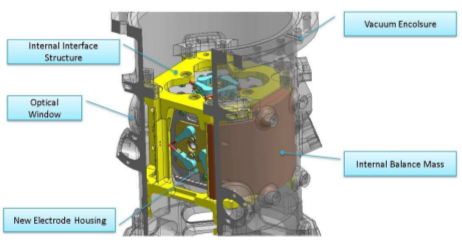} }%
\qquad
\subfloat[Shape and features.]{\includegraphics[width=150pt]{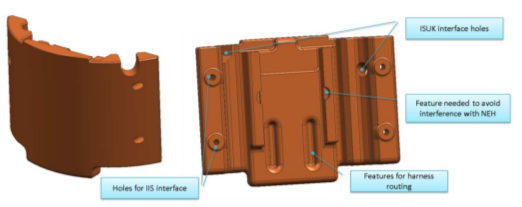}}%
\caption{Internal balance mass.}
\label{f5}
\end{center}
\end{figure}

\begin{figure}[htbp]
\begin{center}
\subfloat[External balance masses shapes and locations.]{\includegraphics[width=150pt]{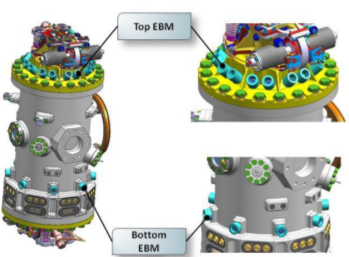}}
\qquad
\subfloat[Spacecraft balance masses shapes and locations.]{\includegraphics[width=170pt]{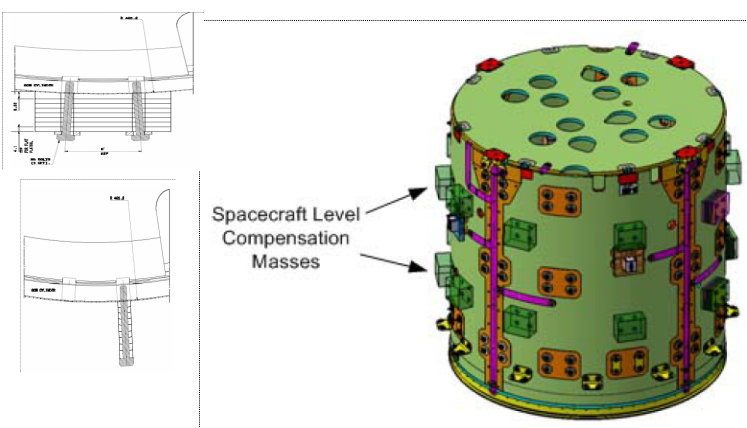}}
\caption{External and S/C balance masses.}
\label{f6f7}
\end{center}
\end{figure}

The only way LPF could comply with the requirements set for the gravitational parameters, was to add some compensating masses. The design of LPF changed repeatedly over the years. In addition, manufacturing and integration processes modified the mass distribution over the time until the construction of the satellite completed. For these reasons, we expected to correct the gravitational imbalance at different stages. We foresaw three sets of balance masses, each one having a certain potential of compensation: the IBMs and EBMs inside LTP, and the spacecraft balance masses (SBMs) outside the LTP's outer cylinder.  Their nominal shapes and locations are shown in Figure \ref{f5} and \ref{f6f7}. The first ones, the IBMs, were designed at the latest stage of the ISH integration and had the capability to compensate the most due to their proximity to the TMs. The EBMs were designed in a second stage, to balance the residual field which was calculated at the end of the LTP assembling. Eventually, we had the SBMs to compensate any residual imbalance after the integration of LTP with the satellite, and, in addition to gravitational compensation, to balance the centre of mass of the spacecraft.  

The IBMs were in the first place nominally designed to compensate a maximum of {\it DC} imbalance of about $50~\mathrm{nms^{-2}}$ with ten $\mathrm{pms^{-2}}$ accuracy. Once the ISH integration terminated, we modified the IBM nominal geometry to adjust the compensation to the up to date imbalance: the procedure was to use circular and rectangular holes, which are in principle easy to machine. In the end, a couple of $\approx 2~\mathrm{kg}$ IBMs compensated an imbalance of $\Delta \vec{a}_{\left(IBM\right)}= [50.31, 0.81, 0.65]~\mathrm{nms^{-2}}$.

The EBMs preliminary design allowed for a maximum of $N=25$ items (for a total EBM mass $<1~\mathrm{kg}$) and a compensation capability of a few tenths of $\,\mathrm{nm/s^2}$. At LTP closure, we used only a few EBMs per ISH,  in total less than $200~g$ mass, to compensate an imbalance of $\Delta \vec{a}_{\left(EBM\right)}= [0.16, 0.13, -0.06]\,\mathrm{nm/s^2}$. 
Only $0.8 \;kg$ of SBMs were needed after the S/C integration with LTP. The final estimate for the fully compensated TM differential gravitational acceleration is:
\be
\label{gravres}
\Delta \vec{a}= \left[ \Delta a_x, \Delta a_y, \Delta a_z \right] = \left[ 0.012,  -0.009,  -0.014 \right]~\mathrm{nm/s^2}\;.
\ee
It is worthy to note that the balancing results in eqn. \ref{gravres} apply to a specific day of LPF mission, that is when the cold gas thruster tanks are full and the DRS 
fuel tanks are half-full. Balancing for that particular date, allowed us to halve the effect of colloidal thrusters on the gravitational imbalance of LPF. The fuel depletion being deterministic, at the beginning of life we will have an additional acceleration of $-0.086~\mathrm{nm/s^2}$ along x due to the extra colloidal propellant. Once all the propellant has been used, the field will have changed to $+0.086~\mathrm{nm/s^2}$.  Instead Cold-gas tanks layout was designed to have a neutral impact on the imbalance, so, in principle, the fuel can be depleted alternatively from different tanks to maintain the imbalance unchanged within $+0.010~\mathrm{nm/s^2}$. 
 
\section{Error budget}
\label{s5}

Although, it is possible to precisely predict the gravitational field generated between the satellite and every TM with the CAD/CAM technology, by design the employment of all the state-of-the-art techniques allows to reach a self-gravitation mass balancing of the order of $10^{-12}\,\mathrm{m/s^2}$ . Unfortunately, current assembling and manufacturing procedures prevent to constrain accuracy at this remarkable level.

The contributions that originate from uncorrected manufacturing errors and the AIT measurement errors are the most important. In this section the error estimates are intended as $3\sigma$ values.

The IBMs inhomogeneities, and defects of fabrication, which turned out to be non detectable during the AIT phase, contribute  $0.115\,\mathrm{nm s^{-2}}$, for the x-acceleration, to the total uncertainty. The reason of this is due to the composition (W($95\%$)) of the IBM, its U-shaped geometry wrapped around the TM, and its proximity to the TM (only a few tens of mm away). To assess that CGS performed gradient density measurements on samples of the same material to estimate worst cases for the IBMs.

As said before in Sect. \ref{s4}, the measurements from AIT propagates their uncertainty to the final results.
The uncertainty on the acceleration due to the uncertainty on the item mass and positioning resulted to be $0.139\,\mathrm{nm s^{-2}}$ and $0.092\,\mathrm{nm s^{-2}}$, respectively, for the $x$-axis acceleration. All these error figure together with the other, minor, ones (LTP and S/C related, not discussed in this paper) summed-up as:
\be
\label{graverrres}
\delta\left(\Delta \vec{a}\right)= [0.283, 0.226, 0.251]~\mathrm{nm/s^2}\;.
\ee
We added the uncertainty terms in the root-mean-square sense in the case of uncorrelated sources, and linearly those from correlated ones. For example we consider uncorrelated the uncertainties originating from mass and positioning measurement errors; on the other hand, the uncertainty originated by the positioning error of a subsystem is the linear sum of that coming from each piece which constitutes the subsystem.

The numbers in the Eqns.~\ref{gravres} and~\ref{graverrres} combine together to produce our estimates with errors, of the DC differential gravitational acceleration on the TMs:

\bea
\label{estdc}
\fl \qquad \left< \Delta {a_x}\right>_{e} = 0.012 \pm 0.283\,\mathrm{nm/s^2}\;;\quad \left< \Delta {a_y}\right>_{e} = -0.009 \pm 0.226\,\mathrm{nm/s^2}\;;\\
\qquad \left< \Delta {a_z}\right>_{e} = -0.014 \pm 0.251\,\mathrm{nm/s^2}\;.\;\nonumber
\eea
These estimates are compliant with the requirements in eqn. \ref{reqdc} and will be verified with the on-board measurements.
In the Appendix we list them together with the values of all the other gravitational parameters. 

\section{Foreseen on-board experimental confirmation technique}
\label{s6}

LPF will release the two TMs from the launch-lock mechanism a few days before entering the science phase, and, at that point, the first experimental data of LPF self-gravity will become available. The on-board measurement of the imbalance between the TMs will be one of the deliverables of noise-run during which we just measure the noise of the instrument; according to Newton's law, the DC acceleration between the TMs, along x-direction is calculated as:
\be
\label{dgmeas}
\Delta A_x = \ddot o_{12} - C_{SUS} \cdot g_c\;.
\ee 

The eqn. \ref{dgmeas} shows that the DC differential acceleration is equal to the double derivative of the mutual position of the TMs ($o_{12}$) minus the control forces per unit mass ($g_c$) applied following the calibration ($C_{SUS}$). 
The DC differential acceleration will contain indistinguishable the gravitational imbalance and other DC forces along x and will be estimated as the time-averaged value of $\Delta A_x$ from eqn. \ref{dgmeas}. 

The accuracy of the measurement is limited by the non-gravitational forces which indistinguishably add-up to the gravitational one. Our present knowledge indicates $0.010~\mathrm{nm/s^2}$ measurement accuracy,  well enough to assess the field within the uncertainties of the analysis and integration.
We will refine our estimate of the other forces during the mission: the LPF experiments would define, little by little, the actual calibration and also reveal the presence of non-gravitational DC terms to improve the accuracy of the measurement.

\section{Conclusions}
\label{s7}

The unprecedented requirements on the self-gravity field on board LPF demanded a new way of designing, manufacturing and assembling a space mission. The procedure put into effects allowed to calculate with enough accuracy all the needed gravitational parameters and to project the balance masses for the needed compensation. 
As soon as LTP operates we are able to verify the results of our calculations with the real picture on-board. In principle, we would be in the condition to measure not only the TM imbalance but also the "angular" imbalance and the self-gravity gradients.
LPF flying will be the test bench to validate the technique of precisely designing the self-gravity of a spacecraft without the need of on-ground experimental apparatus. 
The success of LPF would pave the way to next generation of drag-free spacecrafts for science mission starting from the satellites for LISA and LISA-like mission. 

\ack
The author acknowledges and thanks ASI and CGS for supporting this work.

\appendix
\section*{Appendix}
In the main text we discussed the gravitational imbalance and its importance in the design and functioning of LPF. As said in the introduction of this paper, the other gravitational parameters have an important impact on the design of the instrument and were subjected to requirements. The verification, on-board LPF, of these requirements is an important legacy to future geodesic space science and for LISA-like mission \cite{eLISA}. In Table \ref{req_others}, we list the estimates and the ranges of tolerable values for all these gravitational parameters. 
They are the unit of mass force, the unit of inertia torque, the linear parasitic gradients of the unit of mass force with respect to the source coordinates, the angular parasitic gradients of the unit of inertia torque with respect to the source rotations, and, finally, the cross gradients, the derivatives of the unit of inertia torque with respect to the source coordinates.
For completeness, we also list in the Table \ref{req_others} the imbalance.

The main message of Table \ref{req_others}, is that all the requirements were met at the final delivery of the hardware with a good margin. 

{\tabulinesep=1.2mm

\begin{table}[hb]
\begin{center}
    \begin{tabu}{ | c | c | c | c | c| c|}
    \hline
    Physical quantity & Parameter & \multicolumn{2}{|c|}{Requirement range} & TM1 estimates & TM2 estimates  \\ \hline
    \multirow{3}{*}{{\bf Differential acceleration [$nms^{-2}$]}} & ${\bf \Delta a_x}$ & {\bf -0.650} & {\bf+0.650 }& \multicolumn{2}{|c|} {{\bf 0.012}} \\ 
    & ${\bf \Delta a_y}$ & {\bf -1.100} & {\bf +1.100} & \multicolumn{2}{|c|} {{\bf -0.009}}  \\ 
    & ${\bf \Delta a_z}$ & {\bf -1.850} & {\bf +1.850} & \multicolumn{2}{|c|} {{\bf -0.014}}  \\ \hline
    \multirow{3}{*}{Linear acceleration [$nms^{-2}$]} & $a_x$ & -10 & +10 & 0.2 & 0.2 \\ 
    & $a_y$ & -10 & +10 & 1.5 & 1.5  \\
    & $a_z$ & -35 & +10 & -4.9 & -4.9  \\ \hline
    \multirow{3}{*}{Angular acceleration [$s^{-2}$]$\times 10^{-9}$} & $\alpha_{\theta}$ & -13.5 & +13.5 & -0.4 & 0.4 \\ 
    & $\alpha_{\eta}$ & -11.5 & +11.5 & 3.1 & -1.2 \\ 
    & $\alpha_{\phi}$ & -8 & +8 & 0.8 & -0.2 \\ \hline 
    \multirow{6}{*}{Linear gradients [$s^{-2}$]$\times 10^{-9}$} & $\Gamma_{xx}$ & -800 & +500  & 354.4 & 361.1 \\ 
    & $\Gamma_{yy}$ & -800 & +500  & -138.2 & -142.3 \\ 
    & $\Gamma_{zz}$ & -1600 & +1000  & -216.1 & -218.7\\ 
    & $\Gamma_{xy}$ & -135 & +135 & 35.5 & 19.6 \\ 
    & $\Gamma_{xz}\;$ & -135 & +135 & 48.0 & -23.0  \\ 
    & $\Gamma_{yz}\;$ & -135 & +135 & -7.6 & 2.6  \\ \hline
  \multirow{9}{*}{Angular gradients [$s^{-2}$]$\times 10^{-9}$} & $\Gamma_{\theta \theta}$ & -2750 & +1960 & 177.2 & 181.0\\ 
    & $\Gamma_{\eta \eta}$ & -2750 & +1960 & 235.0 & 231.2 \\     
    & $\Gamma_{\phi \phi}$ & -2070 & +1480 & 140.8 & 143.5  \\     
    & $\Gamma_{\theta \eta}$ & -60 & +60 & -2.7 & -1.0  \\ 
    & $\Gamma_{\theta \phi}$ & -60 & +60 & -4.8 & 2.0 \\ 
    & $\Gamma_{\phi \theta}$ & -60 & +60 & -1.8 & 0.8  \\     
    & $\Gamma_{\eta \phi}$ & -60 & +60 & 0.8 & -0.8  \\ 
    & $\Gamma_{\eta \theta }$ & -60 & +60 & -3.5 & -0.8 \\ 
    & $\Gamma_{\phi \eta}$ & -60 & +60 & 1.3 & -1.2 \\  \hline
    \multirow{9}{*}{Cross gradients [$m^{-1}s^{-2}$]$\times 10^{-9}$} & $\Gamma_{\theta x}$ & -1205 & +1205 & -36.8 & -36.9  \\ 
    & $\Gamma_{\theta y}$ & -1205 & +1205 & 199.4 & 218.7 \\     
    & $\Gamma_{\theta z}$ & -1205 & +1205 & 69.2 & 2.8  \\     
    & $\Gamma_{\eta x}$ & -1205 & +1205 & -179.1 & -287.8 \\ 
    & $\Gamma_{\eta y}$ & -1205 & +1205 & 28.5 & 27.8 \\ 
    & $\Gamma_{\eta z}$ & -1205 & +1205 & -495.7 & 412.1 \\     
    & $\Gamma_{\phi x}$ & -1205 & +1205 & 104.5 & 4.02\\     
    & $\Gamma_{\phi y}$ & -1205 & +1205 & -436.4 & 311.4 \\ 
    & $\Gamma_{\phi z}$ & -1205 & +1205 & 8.3 & 9.1 \\  \hline
    \end{tabu}
\end{center}
\caption{The gravitational parameter requirements and calculated values.}
\label{req_others}
\end{table}

\end{document}